\documentclass{ngsm2024} 
\pdfoutput=1

\usepackage{todonotes}
\usepackage{color}

\title{ECG Signal Denoising Using Multi-scale Patch Embedding and Transformers}

\optauthor{%
\Name{Ding Zhu} \Email{zhu.3723@osu.edu}\\
\Name{Vishnu {Kabir Chhabra}} \Email{chhabra.67@osu.edu}\\
\Name{Mohammad {Mahdi Khalili}} \Email{khalili.17@osu.edu}\\
\addr Ohio State University, US}

\begin{document}

\maketitle

\begin{abstract}%
    Cardiovascular disease is a major life-threatening condition that is commonly monitored using electrocardiogram (ECG) signals. However, these signals are often contaminated by various types of noise at different intensities, significantly interfering with downstream tasks. Therefore, denoising ECG signals and increasing the signal-to-noise ratio is crucial for cardiovascular monitoring. In this paper, we propose a deep learning method that combines one-dimensional convolutional layers with Transformer architecture for denoising ECG signals. The convolutional layers process the ECG signal by various kernel/patch sizes and generate an embedding called multi-scale patch embedding. The embedding then is used as the input of a Transformer-based network and enhances the capability of the Transformer for denoising the ECG signal.
\end{abstract}


\section{Introduction}

The electrocardiogram (ECG) signal is a readily accessible physiological signal that can be used to assess human health \cite{9995675} and monitor cardiovascular disease \cite{hannun2019cardiologist}. With the advancements in wearable sensors and devices, acquiring daily ECG signals has become simple and feasible. However, with the increasing variety of data acquisition scenarios, the noise characteristics in the acquired ECG signals have evolved, differing from those observed in the past. This can make subsequent health monitoring tasks difficult. 

Numerous methods exist for reducing noise in ECG signals. Traditional approaches primarily employ fixed or adaptive filtering for denoising \cite{9210790, martis2013ecg}. However, these methods often focus on the characteristics of specific frequencies, such as high or low frequencies. Recent research has demonstrated promising results using deep learning methods for image denoising, and these approaches have been extended to denoise ECG signals as well. Most of these approaches are based on Denoising Autoencoders (DAEs), which transform the ECG signals into high-level feature representations before reconstructing the denoised signals from these features. While there have been numerous studies on deep learning-based denoising of ECG signals, many of these methods are tailored to specific types of noise and may not perform well on signals with low signal-to-noise ratios. In particular, the following noise types exhibit different frequencies and intensities, and we need a method that can capture each noise component,  \textbf{Baseline Wander}:  This type of noise is caused mainly by respiration or body movement and is usually present in the original signal in a low-frequency form. \textbf{Powerline Interference}: This type of noise is caused by inductive and capacitive couplings of 50/60 Hz power lines during ECG signal acquisition.  \textbf{Electrode Contact Noise}: These noises are caused by improper contact between the body and electrodes.
\textbf{Muscle Contraction}: This type of noise is caused by electrical activities in muscles or other tissues of the human body such as the physiological electrical signals generated by muscle contraction. This type of noise has a frequency ranging from 0.01 Hz to 100 Hz and it tends to distort local waves of the ECG signals.

In this paper, we introduce a new method of time series denoising and conduct experiments on ECG data. Our method combines one-dimensional convolutional layers with Transformer architecture. The convolutional layers processe the time series data and generate embedding used as the input of the Transformer network.  We will show that if the convolutional layer uses different kernel sizes, it can capture different types of noise and artifacts and enable the Transformer to better denoise the ECG signals compared to existing baselines.

\noindent\textbf{Related Work.} Conventional ECG noise reduction methods tend to be performed using filtering, generally utilizing a thresholding approach for certain signals. These thresholding methods include the Fourier transform, Principal Component Analysis (PCA) \cite{bera2019preserving}, Empirical Mode Decomposition (EMD) \cite{kabir2012denoising}, Discrete Wavelet Transform (DWT) \cite{martis2013ecg}. Although these methods can partly reduce ECG noise, hyperparameter tuning requires significant domain knowledge \cite{kabir2012denoising}, furthermore, these methods cannot reduce various types of noise in tandem. Additionally, a significant challenge with these methods is effectively partitioning the signal into sub-signals to ensure that the original signal's information is preserved during the thresholding process without losing information.

Moreover, expeditious development of deep learning methods \cite{vaswani2017attention,kingma2022autoencoding,gu2023mamba,beck2024xlstm} has resulted in the emergence of promising solutions to ECG noise reduction\cite{zhao2021hybrid}. Some of these methods are based on the autoencoder (AE) architecture, which can regenerate the original signal without requiring extensive domain knowledge \cite{zhao2021hybrid}. Poungponsri and Yu in \cite{poungponsri2009electrocardiogram} proposed a wavelet neural network for ECG noise reduction. This work first decomposed the signal into multiple components using wavelets and then reconstructed the signal in a six-layer convolutional encoder-decoder network. Antczak \cite{antczak2018deep} proposed a Deep Recurrent Neural Network (DRNN) to denoise ECG signals. The model is a hybrid of DRNN and a denoising autoencoder, which is trained to recreate noisy input data and achieved better results than traditional methods. Qiu et al. \cite{qiu2020two} presented a two-stage denoising method for removing noise from ECG signals. It included an improved one-dimensional U-net, which enhances the size of the convolution kernel, and a DR-net for detailed restoration in the second stage. He et al. \cite{he2021dual} proposed a dual attention convolutional neural network based on adaptive parametric ReLU for denoising ECG signals with strong noise. The network is designed with a dual attention module and a modified activation function, resulting in significant noise reduction. Singh and Sharma \cite{9853604} introduced a novel attention-based convolutional denoising autoencoder (ACDAE) model that employs an attention module and skip-layer to reconstruct the ECG signal. Additionally, the intermediate layer features are utilized to train a classifier in this study.

\section{Methodology}
\begin{figure}[h]
\centering\includegraphics[width=0.8\textwidth]{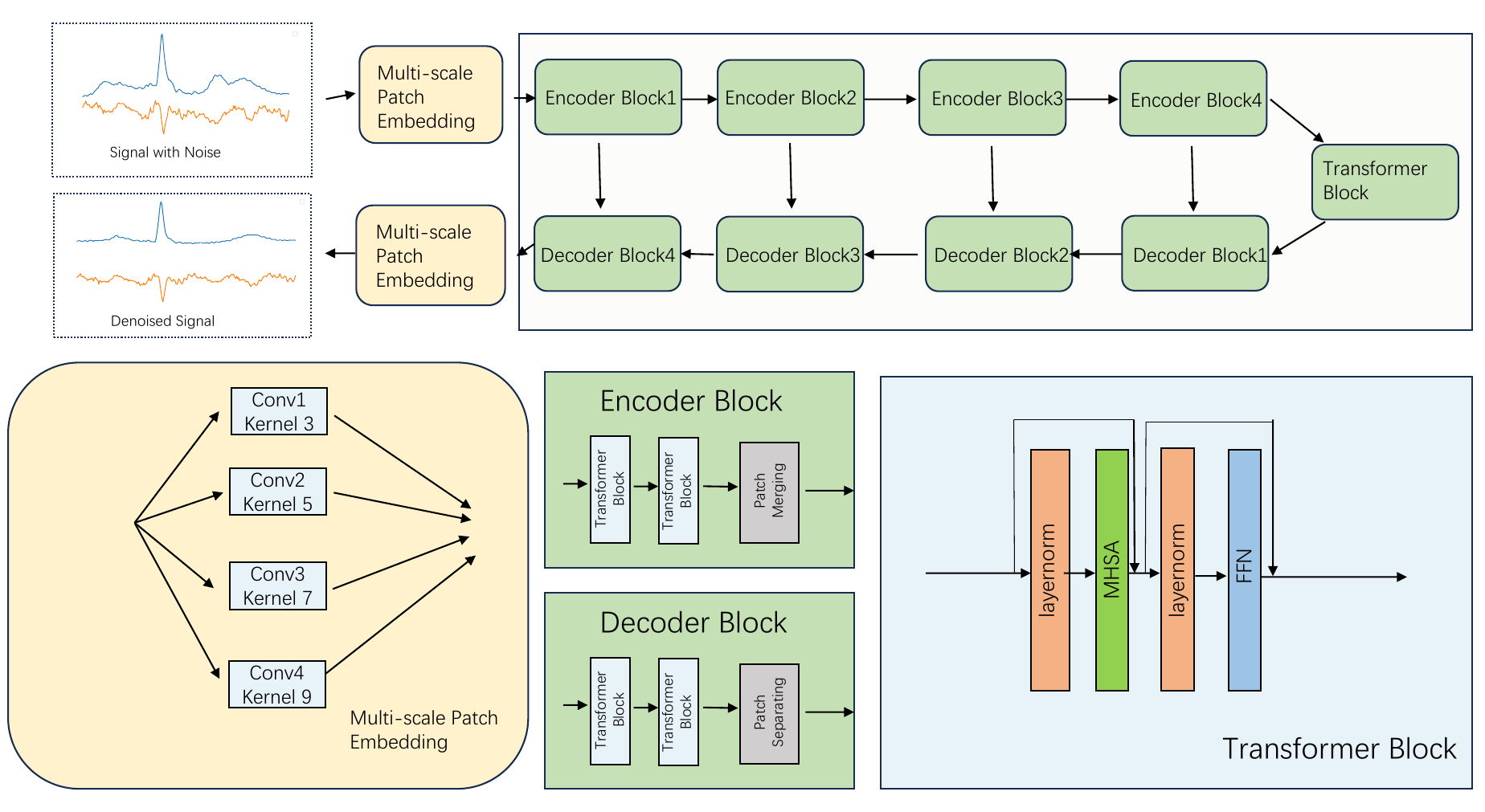}
    \caption{Proposed architecture for denoising. Each encoder block consists of two Transformer blocks and a patch merging module. The decoder block consists of two Transformer blocks and a patch separating module. Each Transformer block consists of two layer normalization, a Multi-Head Self-Attention (MHSA), and a Feed Forward Network (FFN).  The embedding layer consists of four convolutional operations with different kernel sizes.}
\label{fig:model_arch}
\end{figure}

\textbf{Overall Architecture.} To denoise time-series signals, our proposed model is based on a U-shaped \cite{ronneberger2015u} network architecture comprising an encoder, a decoder, and two convolutional layers (one before the encoder and one after the decoder). The encoder and decoder are built based on the Transformer architecture and include several skip connections. The time series data first passes through a convolutional layer with different kernel sizes to generate embedding (we call this layer a multi-scale patch embedding layer). The embedding then will pass through the encoder and decoder which contain several Transformer blocks. Between these blocks, patch merging and separating are used to operate on features to reshape them. 
After the decoder, there is another convolutional layer (i.e., Multi-scale patch embedding layer) to reconstruct the time-series data. Figure \ref{fig:model_arch} illustrates the proposed architecture. We use Mean Squared Error as the loss function to train the proposed model.

\noindent\textbf{Multi-scale Patch Embedding Layer.} Frequency is an important factor in time series, especially in electrocardiogram (ECG) signals, where different regions exhibit distinct frequency characteristics. In ECG signals, different regions of the heart exhibit unique frequency characteristics, which can be critical for accurate diagnosing and monitoring. These distinct frequency features can help identify various cardiac conditions, enabling more effective and targeted medical interventions. Understanding and analyzing these frequency variations within ECG signals is essential for healthcare professionals to make informed decisions regarding patient care. We propose utilizing a multi-scale patch embedding layer to capture these distinct frequency characteristics. This layer employs convolutional operation with different kernel/patch sizes (e.g., in our experiment, we use kernel sizes 3, 5, 7, 9). The choice of patch sizes depends on the temporal granularity we aim to capture. The parameters of the embedding layer can be initialized randomly or using pre-trained kernels. The embeddings from patches of different sizes are concatenated to create a multi-scale representation. This approach enables the model to leverage information from multiple temporal resolutions. The multi-scale representation then can be used as an input for the encoder. We   use a similar multi-scale patch embedding to process the output of the decoder and generate the final denoised signal.  

\noindent\textbf{Patch Merging/Separating.} As shown in Figure \ref{fig:model_arch}, each encoder block includes a patch merging module that reshapes the signal by decreasing the length and increasing the channel size. The decoder block includes a patch separating module which increases the length and decreases the channel size.

\noindent\textbf{Masked Input.} Recent studies show masking the input during the training can improve the model performance \cite{li2023timae}. Masks are binary matrices/tensors and generated randomly for each data point in each iteration during the training. In each iteration, there will be an element-wise product between each data point and its mask before the forward and backward propagation. We use masked input during the training of the proposed network to improve the denoising performance of our model.  


\section{Experiments}
\textbf{Datasets and Settings.}
In our experiment, we utilize the MIT-BIH Arrhythmia Dataset \cite{moody2001impact}, a well-known ECG signal dataset. This dataset comprises 48 two-channel clean ECG recordings from 47 patients, with each recording containing 650,000 samples at a sampling frequency of 360Hz (approximately half an hour of data). This data serves as the predicted output for our model. To simulate real-world conditions, we manually added noise to the dataset using real ECG noise from the MIT-BIH Stress Test Database. This dataset includes three types of noise: baseline wander, muscle artifact, and electrode motion artifact, represented as 'bw', 'ma', and 'em', respectively. Both datasets are available for download from PhysioNet \cite{goldberger2000physiobank}.

Each time series ECG signal is segmented into smaller parts, each containing 256 samples, which corresponds to one heartbeat. We do the same for the MIT-BIH Stress Test Dataset to create noise samples with a length of 256.  To generate a training dataset, we pick a heartbeat segment and a noise segment randomly and add them to generate a noisy input. The noise signal strength is adjusted by a constant to meet the specified signal-to-noise ratio (SNR).  The noisy heartbeat segment will be the input of our network. The original ECG segment is used as the target output.
 
The noise ECG signals for the experiment consist of four distinct classes: the first three represent individual noise types ('bw', 'ma', and 'em'), and the fourth class, 'emb', is a combination of all three noise types. The generated dataset contains 103,091 data points of length 256, which are partitioned into training and testing sets at a ratio of 4:1. 

 We employ two metrics to evaluate the network model: Root Mean Square Error (RMSE) and Signal to Noise Ratio (SNR). For RMSE, a smaller value indicates better performance, while for SNR, a higher value indicates better performance. We provide a description of the hyper-parameters used in our model in the appendix.

\noindent\textbf{Denoising Performance of the Proposed Model.}  We perform denoising experiments on different types of noise, specifically baseline wander (bw), electrode motion (em), muscle artifact (ma), and a combination of these three types, all at a noise intensity of -4dB. \vspace{-0.35cm}

\begin{table*}[htbp]  \caption{Performance under different types of noise}
    \begin{center}
    \begin{tabular}{ccc ccc ccc}
    \hline
    \textbf{Methods}&\multicolumn{4}{c}{\textbf{Signal to Noise Ratio(dB)}} &\multicolumn{4}{c}{\textbf{Root Mean Squared Error}}\\
    \cline{2-9} 
    \textbf{} & \textbf{\textit{bw}}& \textbf{\textit{em}}& \textbf{\textit{ma}} &\textbf{\textit{ebm}} &\textbf{\textit{bw}}& \textbf{\textit{em}}& \textbf{\textit{ma}} &\textbf{\textit{ebm}} \\
    \hline
    DWT     & 1.00    & 1.25    & 1.32& 1.23 & 1.58& 1.55& 1.58 &1.58\\

    U-Net\cite{ronneberger2015u}   & 10.71  & 7.90  & 8.16  & 7.82  &       0.32   & 0.42            & 0.41            & 0.42\\

    DACNN\cite{he2021dual}           & 13.1  & 9.46  & 9.30  & 9.18  & 0.25           & 0.36           & 0.36            & 0.37\\
    \hline
    Ours & 14.84  & 12.75 &12.11  & 12.62  & 0.21 & 0.26 &0.28 & 0.27\\
    \hline
 
    \end{tabular}
    \label{snr_diff_type}
    \end{center}
\end{table*}

For comparison, we test traditional denoising methods, specifically discrete wavelet transform (DWT) thresholding, as well as U-Net \cite{ronneberger2015u} and DACNN \cite{he2021dual}, alongside our model. For the DWT method, we select 'db8' as the wavelet base and apply soft thresholding. It can be seen in Table \ref{snr_diff_type} that our method achieves the lowest RMSE and the highest SNR under different noise types. Additionally, we can see that all methods perform well under 'bw' noise. This is because 'bw' noise has a relatively single frequency, making it easier for different methods to eliminate.

\section*{Acknowledgement}

This material is based upon work supported by the U.S. National Science Foundation under award  IIS-2301599 and CMMI-2301601, by grants from the Ohio State University's Translational Data Analytics Institute and College of Engineering Strategic Research Initiative.







\bibliography{ngsm}

\section{Appendix}

\paragraph{Hyper-parameters in the Experiment} As shown in Figure \ref{fig:model_arch}, we use an architecture with four encoder and decoder blocks. The input size of multi-scale patch embedding is 2 by 256 (there are two channels. The length of signal in each channel is 256). The size of the generated embedding is 8 by 256. We have the following hyper-parameters for the encoder and decoder blocks, Multi-Head Self-Attention (MHSA) modules in Encoder Block 1 and Decoder Block 4 have an embedding size of $8$ and include two heads. MHSA modules in Encoder Block 2 and Decoder Block 3 have an embedding size of 16 and include four heads. MHSA modules in Encoder Block 3 and Decoder Block 2 have an embedding size of 32 and include eight heads. Lastly, MHSA modules in Encoder Block 4 and Decoder Block 1 have an embedding size of 64 and include 16 heads. The patch merging module would double the number of channels and reduce the length by half, and the patch separating module double the length and decrease the the number of channels by half. The learning rate is 0.001 and we trained for 100 epochs.

\paragraph{Classification of the Denoised Data}
To measure the impact of denoised signals on downstream tasks, in this part, we use the denoised ECG signals for classification. In this experiment, these signals are contaminated with mixed 'em', 'bw', and 'ma' noise at a strength of -4 SNR.

We train a ResNet network using the clean ECG data and then use the denoised signals to make predictions through the network.  To create a dataset, we randomly select 7,000 samples of N-type ECG data representing normal beats, and 7,000 samples of V-type ECG data representing abnormal beats.  The label indicates the type of ECG data (N-type and V-type). We pick 5,000 samples of each category for the training set, while 2,000 samples of each category are set aside as the test set. The classifier is trained using clean training data, comprising a total of 10,000 samples. The training process for the classifier was conducted over 20 epochs. 

In table \ref{acc_compare}, the first line includes the performance of the classifier on the clean test data, the second line includes the performance of the classifier on noisy test data, and the other lines include the performance of the classifier after denoising the input data.   

\begin{table}[htbp]
  \caption{Classification experiments based on different denoising methods.}
  \begin{center}
  \begin{tabular}{cccc}
  \hline

  \textbf{Methods} & \textbf{\textit{accuracy}}& \textbf{\textit{precision}}& \textbf{\textit{f1-score}} \\
  \hline
  original data & 0.809 & 0.916 & 0.781\\
  \hline
  NOP   & 0.599          & 0.642          & 0.528          \\
  DWT   & 0.597          & 0.639          & 0.527          \\
  U-Net & 0.684          & 0.827          & 0.596          \\
  DACNN & 0.672          & 0.819          & 0.573          \\
  \textbf{Ours}  & \textbf{0.794} & \textbf{0.886} & \textbf{0.766} \\
  \hline
  \end{tabular}
  \label{acc_compare}
  \end{center}
\end{table}

Table \ref{acc_compare} shows that our denoising approach can lead to better classification accuracy compared to baselines.

\end{document}